\documentclass[aps,prl,twocolumn,superscriptaddress,showpacs,amsmath,amssymb]{revtex4}

\usepackage{graphicx}

\begin{document}

\title{Dynamics of matter-wave solitons in a ratchet potential}

\author{Dario Poletti}

\affiliation{Department of Physics and Centre for Computational
Science and Engineering, National University of Singapore, Singapore
117542, Republic of Singapore}

\affiliation{ARC Centre of Excellence for Quantum-Atom Optics and
Nonlinear Physics Centre, The Australian National University,
Canberra ACT 0200, Australia}

\author{Tristram Alexander}

\affiliation{ARC Centre of Excellence for Quantum-Atom Optics and
Nonlinear Physics Centre, The Australian National University,
Canberra ACT 0200, Australia}

\author{Elena Ostrovskaya}

\affiliation{ARC Centre of Excellence for Quantum-Atom Optics and
Nonlinear Physics Centre, The Australian National University,
Canberra ACT 0200, Australia}

\author{Baowen Li}

\affiliation{Department of Physics and Centre for Computational
Science and Engineering, National University of Singapore, Singapore
117542, Republic of Singapore} \affiliation{NUS Graduate School for
Integrative Sciences and Engineering, 117597, Republic of Singapore}

\author{Yuri S. Kivshar}

\affiliation{ARC Centre of Excellence for Quantum-Atom Optics and
Nonlinear Physics Centre, The Australian National University,
Canberra ACT 0200, Australia}

\date{\today}

\begin{abstract}
We study the dynamics of bright solitons formed in a Bose-Einstein
condensate with attractive atomic interactions perturbed by a weak bichromatic optical
lattice potential. The lattice depth is a biperiodic function
of time with a zero mean, which realises a flashing
ratchet for matter-wave solitons. The average velocity of a soliton and the directed soliton current induced by the ratchet depend on the number of atoms in the soliton. We employ
this feature to study collisions between ratchet-driven solitons and find that soliton transport can be induced
through their interactions. In the regime when matter-wave solitons
are narrow compared to the lattice period the ratchet dynamics is
well described by the effective Hamiltonian theory.
\end{abstract}


\maketitle

The ratchet effect, i.e. rectified average current induced by an
asymmetric potential and unbiased zero-mean driving, has
historically attracted a lot of attention due to its possible
relevance to biological transport, and molecular motors, and prospects
for nanotechnology~\cite{Hanggi,Reimann}. Both in classical and quantum systems, the
ratchet effect has been studied both in dissipative  and Hamiltonian
regimes~\cite{Hanggi2, Monteiro} and it appears to be due to a broken space-time symmetry of
the perturbing potential~\cite{Flach}. Directed transport can be experimentally implemented in a range of
different physical systems ranging from semiconductor
hetero-structures to quantum dots, Josephson junctions, and cold
atoms in optical lattices~\cite{Renzoni}. Recently, the interest in
optical ratchets and especially in the effect of interaction on the
ratchet transport~\cite{Dario,Luis,Romero} has resurged with the
experimental advances in implementing atomic ratchets for
Bose-Einstein condensates~\cite{BEC_exp1,BEC_exp2}.

As a physical system with intrinsically present nonlinear
interactions, Bose-Einstein condensates (BECs) of atoms with negative
scattering length support the existence of localized collective
excitations - bright matter-wave solitons. It is therefore
interesting to explore the effect of the ratchet potentials on the
transport and interaction properties of such collective
excitations. It is especially important because theoretical
studies of the soliton ratchets so far were mostly focused on topological
solitons~\cite{Kink}, whereas BECs represent a perfect test-bed for
the study of the ratchet dynamics of a general class of
non-topological solitons described by a continuous Gross-Pitaevskii
(or nonlinear Schr\"odinger) equation.

In this Letter, we study, for the first time to our knowledge, the
effect of the ratchet potential on non-topological non-dissipative
bright matter-wave solitons. A BEC represents an intrinsically
nonlinear system which displays a single-particle behavior on a
macroscopic scale being easily manipulated by reconfigurable optical
and magnetic potentials.  The ratchet potential is realized by means
of a bichromatic optical lattice which is ``flashed'' on and off in
such a way that its time-averaged amplitude vanishes. We show that
both the ratchet effect and soliton directed transport are
observed in this system even in the absence of losses, which sets it
apart from previously studied dissipative nonlinear systems
subjected to ratchet potentials~\cite{FlachOpt}. A weak potential
does not affect the soliton shape in a wide range of parameters,
and especially in the regime when the solitons are strongly
localized, which justifies treatment of the soliton as an effective
classical particle. However, we show that the ratchet still works
even when the extended nature of the excitation cannot be ignored
and the soliton width exceeds the period of the lattice. Moreover,
we investigate the influence of the ratchet on soliton scattering,
and show that multiple collisions between solitons may provide a
space averaging mechanism that can enable observation of a soliton
current in a ratchet potential.

To study the dynamics of bright solitons in attractive BECs with
ratchet potentials we consider a soliton formed in a strongly
elongated condensate cloud~\cite{BEC_soliton} subjected to a flashing
one-dimensional optical lattice. As long as the energy of the
longitudinal excitations is not sufficient to excite the transverse
modes of the condensate, the system can be treated as
one-dimensional and described by the Gross-Pitaevskii (GP) equation:
\begin{equation}\label{GP1D}
i\frac {\partial \Psi}  {\partial t}+\frac{1}{2}\frac{\partial^2
\Psi}{\partial x^2}+|\Psi|^2\Psi- V(x,t)\Psi=0,
\end{equation}
where the Fourier-synthesized lattice potential,
\begin{equation}\label{pot}
V(x,t)=V_{0}\; f(t)\left[\cos(x)+\cos(2x+\phi)\right],
\end{equation}
is driven biperiodically: $f(t)=\sin(\omega t)+\sin(2 \omega
t )$, and $V_0$ depends on the intensity of the laser beams forming
the lattice.  Quantum transport of atoms in the stationary potential
of the form (\ref{pot}) has been recently studied experimentally
with ultracold rubidium vapor~\cite{Weitz}. With the time-dependent
potential (\ref{pot}) the space inversion symmetry is broken for
$\phi\ne 0,\pi$ and the time inversion symmetry is broken by our
choice of $f(t)$. According to the symmetry analysis~\cite{Flach},
this can allow a directed transport of matter-wave solitons.

In derivation of the dimensionless model~(\ref{GP1D}) we have
assumed that the energy, length, and frequency are measured in the
units of $E_0=\hbar^2k^2/m$, $a_0=1/k$, and $\omega_0=\hbar k^2/m$,
respectively, where $m$ is the mass of the atoms, and $k$ is the
wavevector of the optical lattice. We have furthermore reduced the
three-dimensional mean-field model to the one-dimensional GP
equation by assuming that the wavefunction is separable in the
following manner: $\psi_{3D}(x,y,z)=\psi_{1D}(x)\Phi(y,z)$, where
$\Phi(y,z)$ is the ground state wavefunction, normalized to one, of the two-dimensional
harmonic oscillator with the transverse trapping frequency
$\omega_\perp$. With these assumptions, the
wavefunction $\Psi$ in Eq.~(\ref{GP1D}) relates to $\psi_{1D}$ as
follows: $\Psi=\psi_{1D}\sqrt{g_{1D}}$, where
$g_{1D}=2(a_s\omega_\perp)/(a_0\omega_0)$ is the renormalized
interaction coefficient that characterizes s-wave scattering of the
condensate atoms with the scattering length $a_s$. Number of atoms
in the system is defined as: ${\cal N}=N/g_{1D}$, where $N=\int
|\Psi|^2 dx$ is the norm of the dimensionless one-dimensional
wavefunction.

In the experimental setup~\cite{BEC_soliton} a bright soliton forms
in the $^7{\rm Li}$ cloud with a modified scattering length
$a_s\approx -0.21\, {\rm nm}$ trapped in a quasi-one-dimensional
atomic waveguide with $\omega_\perp=2\pi\times 710 \, {\rm Hz}$. We
consider this experimental setup with an additional optical lattice
formed by light beams crossed at the angle $\theta=38^{\circ}$,
derived from a $CO_2$ laser with the wavelength $\lambda=10.62 \,
\mu m$. With these assumptions the scaling units of length and
frequency take the values $a_0=\lambda/[4\pi\sin(\theta/2)]=2.52 \,
\mu m$ and $\omega_0=2\pi \times 224 \, Hz$. A stable
bright soliton typically created in the
experiment~\cite{BEC_soliton} contains ${\cal N}\approx 5\times
10^3$ atoms, which corresponds to $N\approx 2.62$ in our units. We
note that by changing the angle $\theta$, it is possible to achieve
smaller or larger values of $a_0$, and hence of $N$, for the same value of ${\cal
N}$.

\begin{figure}[!h]
\includegraphics[width=6.5cm, keepaspectratio=true]{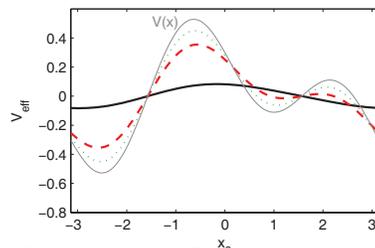}
\vspace{-0.6cm} \caption{(Color online) Effective potential $V_{\rm
eff}$ vs. soliton position $x_0$, for $N=1$ (solid line), $N=3$
(dashed) and $N=5$ (dotted). Thin solid line is the optical
lattice potential $V(x,t)$. Parameters are: $V_0=0.3$, $\phi=\pi/2$,
$f(t)=1$ }
\label{fig:asym}
\end{figure}

At $t=0$, i.e. in the absence of the lattice, the shape of the soliton is
given by the exact solution of the GP equation,
\begin{equation}
\Psi(x,x_0,0)=(N/2) {\rm sech}\left[ (N/2) \left(x-x_0\right)\right],
\end{equation}
where $x_0$ is the initial position of the soliton, and the motion
of the soliton is free. If we assume that the addition of
a weak ratchet potential does not affect the soliton shape during
its evolution, then the Hamiltonian description of the mean field
allows us to treat the matter-wave soliton as an effective classical
particle~\cite{soliton_particle} that moves in the effective
potential,
\begin{eqnarray}\label{eff_pot}
V_{\rm eff}(x_0,t)=\frac 1
N\int_{-\infty}^{\infty}|\Psi(x,x_0,0)|^2V(x,t)dx=\\ \nonumber
=\frac{\pi} N V_0\;f(t) \left[ \frac {\cos x_0}{{\rm
sinh}(\pi/N)}+ 2\frac {\cos (2x_0+\phi)}{{\rm sinh}(2\pi/N)}\right
].
\end{eqnarray}
Using Eq.(\ref{eff_pot}) and the equation of motion for the soliton position, $d^2x_0/dt^2=-dV_{\rm
eff}/dx_0$, the soliton velocity can be obtained straightforwardly. The instantaneous
shape of the effective potential, shown in Fig.~\ref{fig:asym} for
$f(t)=1$, becomes closer to the shape of the optical lattice,
$V(x,t)$ (Fig.~\ref{fig:asym}, thin solid line), as $N$ grows and the
soliton becomes more localized.  For small $N$ the second term in
(\ref{eff_pot}) becomes exponentially smaller than the first one,
and hence the effective potential becomes practically symmetric, as
demonstrated by a solid thick line in Fig.~\ref{fig:asym}.

\begin{figure}[!h]
\includegraphics[width=\columnwidth, keepaspectratio]{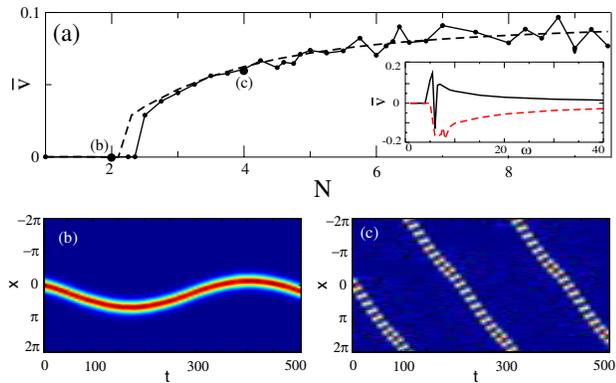}
\vspace{-0.6cm} \caption{(Color online) (a) Cumulative velocity
${\bar v}$ vs. number of atoms in the soliton, $N$, calculated using Eq.(\ref{GP1D}) (solid line) and EPA (dashed); ${\bar v}=1$ corresponds to $3.5$ mm/s. Parameters are: $V=0.3$, $\phi=\pi/2$,
$\omega=10$, $x_0=0$. Inset: Cumulative velocity ${\bar
v}$ vs. driving frequency $\omega$, for $N=4$ and $x_0=0$ (solid line) and $x_0=-\pi/2$ (dashed line). (b,c) Density plot of the mean
field evolution, $|\psi(x,t)|^2$, shown for the corresponding points
at $N=2$ and $N=4$ in (a).} \label{fig:vvsN}
\end{figure}

The effective-particle approximation (EPA) predicts that the
cumulative velocity of the collective excitation, $\bar v=(1/T)\int_0^T v(t) dt$,
is a function of the number of atoms contained in the soliton ($N$).  As can be seen in
Fig.~\ref{fig:vvsN}, this prediction is in good agreement with the
numerical results obtained from the GP model, where we have used 
$T\approx 10^3\times 2\pi/\omega$.  As Fig.~\ref{fig:vvsN}
demonstrates, for a fixed initial position of the soliton, $x_0$,
there is a sharp transition between a regime where a soliton oscillates between neighboring wells but is not
transported [Fig.~\ref{fig:vvsN}(b)], and a regime where the soliton
acquires a ballistic motion [Fig.~\ref{fig:vvsN}(c)]. The
simulations of ballistic motion are performed with the periodic
boundary conditions, however we make sure that the tails of the soliton do not overlap.  The EPA model also predicts that the velocity
of the ballistic motion tends to a constant value as $N\to \infty$
(and $T\to \infty$), which is qualitatively confirmed by numerical
simulations. 

The dependence of the cumulative velocity ${\bar v}$ on
the number of atoms is also a function of the driving frequency,
$\omega$ [see inset in Fig.~(\ref{fig:vvsN})]. For high frequencies
the velocity vanishes as $v\propto\omega^{-1}$, as the soliton
becomes less and less affected by the rapidly oscillating potential.
For the moderate driving frequencies used here the soliton does not
radiate, and small variations of its shape do not lead to particle
loss. In the limit of small $\omega$ the soliton strongly interacts
with the lattice and may break up. In the effective-particle model
the trajectory of the soliton in this case is chaotic and a detailed
study of its behavior is beyond the scope of this paper.

\begin{figure}[!h]
\includegraphics[width=\columnwidth, keepaspectratio]{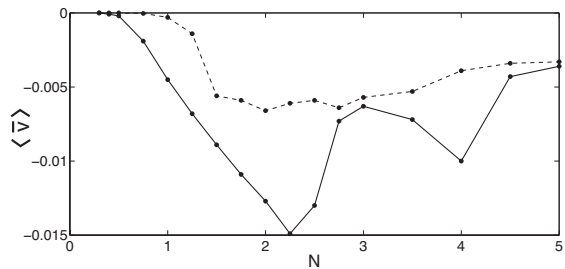}
\vspace{-0.6cm} \caption{(a) Average velocity $\langle {\bar v}
\rangle$ vs. number of atoms in the soliton, $N$, calculated using the GP model (solid line)
and EPA (dashed). Parameters are: $V_0=0.3$,
$\phi=\pi/2$, $\omega=10$.} \label{fig:avvvsN}
\end{figure}

The dependence of the soliton velocity on atom number is a general feature that occurs for symmetric periodic potentials as well,
however the precise relationship between velocity and atom number
depends on the initial position of the soliton relative to the
lattice.  Since the space inversion symmetry of the lattice is
broken, we can expect that averaging over all initial soliton
positions, $x_0$, will lead to a directed soliton current.  To
demonstrate this effect, in Fig. \ref{fig:avvvsN} we plot the
soliton velocity averaged over all initial positions, $\langle {\bar
v} \rangle= (1/2\pi)\int_{0}^{2\pi}{\bar v} dx_0$, as a function of
the number of atoms. Although the average velocity is always
non-zero, one can identify two different regimes of the ratchet
dynamics depending on the number of atoms, as discussed below.

\begin{figure}[!h]
\includegraphics[width=\columnwidth,  keepaspectratio]{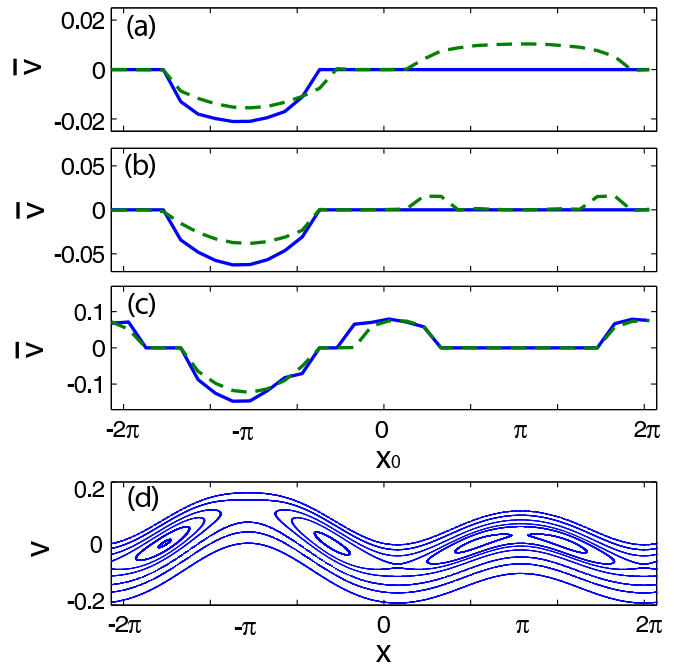}
\vspace{-0.6cm} \caption{(Color online) (a) Cumulative velocity, ${\bar v}$, vs.
soliton initial position, $x_0$, for (a) $N = 1$; (b) $N=2$; (c)
$N=5$; calculated from the numerical solution of the GP equation
(solid line) and EPA (dashed). (d) Poincar\'{e} section for the
effective-particle model in (c). Parameters are: $V_0=0.3$,
$\phi=\pi/2$, $\omega=10$.} \label{fig:NVpos}
\end{figure}

As seen in Fig.~\ref{fig:avvvsN}, for small values of $N$ the EPA
results and numerical soluton of Eq.(\ref{GP1D}) disagree both on the onset of the ratchet
effect and on its magnitude.  For $N<2.5$ the soliton's size is
comparable to or larger than a period of the optical lattice. Hence
it is more accurately described as a wavepacket than an effective particle. The
details of the soliton response to the flashing potential are best
seen by examining the dependence of the cumulative velocity ${\bar
v}$ on the initial position of the center-of-mass, $x_0$.  In
Figs.~\ref{fig:NVpos}(a,b) we show this response for $N = 1$ and
$N=2$. Interestingly the numerical results show that the soliton has either no cumulative
velocity, or moves in only one direction. As a result, for $N<2.5$ a
soliton attains a much larger average velocity than that predicted
by the EPA theory (see Fig.~\ref{fig:avvvsN}). In fact the EPA incorrectly predicts that the
soliton can move in both directions depending on its initial
position. We also note that for $N=1$ the cumulative velocity
expected from the EPA is almost symmetric around $x_0=0$ as expected
in the limit of small $N$ due to the symmetry of the effective
potential (see Fig.\ref{fig:asym}).

For large values of $N$, corresponding to a strongly localized
matter-wave soliton, the ratchet dynamics is well described by the
EPA.  In Fig.~\ref{fig:avvvsN} we observe a good agreement
between the ratchet velocity predicted by the effective particle
model and that found from solving Eq.(\ref{GP1D}) numerically.
Similarly, in Fig.~\ref{fig:NVpos}(c) we see a good agreement
between the numerics and the EPA in the details of the cumulative
velocity dependence on $x_0$.  In
Fig.~\ref{fig:NVpos}(d) we show the Poincar\'e section corresponding
to the effective particle results of Fig.~\ref{fig:NVpos}(c). Two
different types of trajectories are observed, transporting
or non-transporting. A comparison between Figs.~\ref{fig:NVpos}(c and d)
shows a clear correlation between cumulative velocity in
Fig.~\ref{fig:NVpos}(c) and either non-transporting trajectories
(correlating with the zero-velocity) or transporting
trajectories (correlating with either positive or negative velocity).

\begin{figure}[!h]
\includegraphics[width=\columnwidth, keepaspectratio]{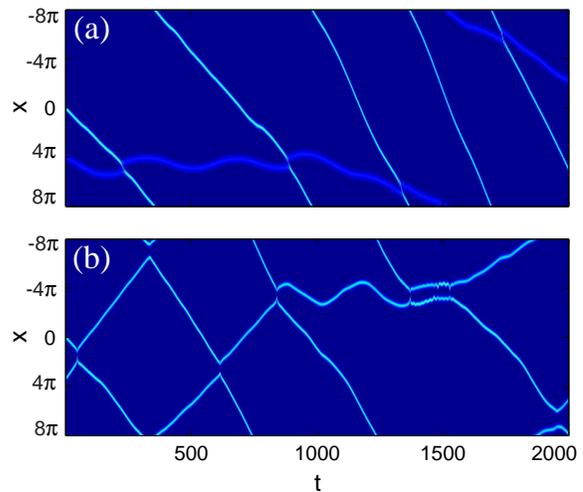}
\vspace{-0.6cm} \caption{(Color online) (a) Collision between two soliton, one with
$N=4$ with initial position $x_0=0$ and the other one with $N=2.2$
and initial position $x_0=4\pi$. (b) Collision between two soliton
with $N=4$ located at $x_0=0$ and $x_0=3\pi+1.2$. Parameters are:
$V_0=0.3$. $\phi=\pi/2$, $\omega=10$.} \label{fig:collisions}
\end{figure}

As shown above,  solitons with different $N$ travel at different ratchet-induced speeds even if their relative initial position in the
lattice is the same.  Multiple soliton collisions could therefore be
realized if the ratchet potential was combined with a toroidal
trap~\cite{Phillips}. As can be seen in Fig.~\ref{fig:collisions}(a)
a larger moving soliton can then induce the transport of a smaller
soliton which would otherwise not be transported by the ratchet.
This is due to the fact that each collision incrementally changes the soliton's
position in the phase space and its original motion can thus be dramatically modified if it is eventually moved from a non-transporting to a
transporting trajectory.  The driving has little effect on the
actual near-elastic soliton collisions because of its small amplitude and fast
variation.

If the solitons have equal values of $N$, collisions can occur only if they have different relative positions in the lattice,
and therefore different speeds.  In this case however, the
interaction between the solitons is strong due to the large
interaction energy.  Each collision therefore induces a much
more pronounced transition to a different phase-space trajectory.  This is evident in
Fig.~\ref{fig:collisions}(b) where we can see dramatic changes in
the soliton velocities after collisions.  In this scenario, the
spatial shift that solitons acquire during each collision may lead
to an effective averaging over $x_0$ after multiple collisions. As a
result, a nonzero total average current can, in principle, be
observed for a sufficiently large number of collisions or for a
sufficiently large number of interacting solitons.

In conclusion, we have studied the ratchet dynamics of bright
solitons in Bose-Einstein condensates and demonstrated that solitons
may either be transported through the potential or oscillate,
depending on their initial position relative to the lattice.  For
small atom numbers, the soliton transport occurs in  one direction
only, while larger solitons may be transported in either direction.
We show that the rate of transport for a given initial position is
atom-number dependent, with solitons that contains more atoms typically moving faster.
Importantly, we find that averaging over all soliton initial
positions reveals an overall directed soliton current, and show
a good qualitative agreement between the numerical results and an effective
particle approximation. Finally, we illustrate how the ratchet potential affects interactions between solitons containing the same or different number of atoms. Collisions can cause an
instantaneous transition between different trajectories in the phase
space, including that between non-transporting and transporting
trajectories. This effect could potentially be used for directed
transport or spatial filtering of solitons based on the atom number. 
This work is supported by the Australian Research Council (ARC).

\end{document}